\documentclass[times,10pt,twocolumn]{article} 
\usepackage{latex8}
\usepackage{times}
\usepackage{graphicx}

\begin{document}

\title{Sound Generation by a Turbulent Flow in Musical Instruments\\
--- Multiphysics Simulation Approach ---}

\author{Taizo Kobayashi$^\dagger$, Toshiya Takami$^\dagger$,
Kin'ya Takahashi$^\ddagger$, Ryota Mibu$^\dagger$,
and Mutsumi Aoyagi$^\dagger$\\ \\
$^\dagger$ Research Institute for Information Technology, Kyushu University\\
6--10--1 Hakozaki, Higashi-ku, Fukuoka 812-8581, Japan\\
\and
$^\ddagger$The Physics Laboratories, Kyushu Institute of Technology\\
Kawazu 680-4, Iizuka 820-8502, Japan\\
}

\maketitle
\thispagestyle{empty}

\begin{abstract}
Total computational costs of scientific simulations are analyzed
between direct numerical simulations (DNS) and multiphysics simulations
(MPS) for sound generation in musical instruments.
In order to produce acoustic sound by a turbulent flow
in a simple recorder-like instrument,
compressible fluid dynamic calculations with a low Mach number
are required around the edges and the resonator of the instrument in DNS,
while incompressible fluid dynamic calculations
coupled with dynamics of sound propagation
based on the Lighthill's acoustic analogy are used in MPS.
These strategies are evaluated
not only from the viewpoint of computational
performances but also from the theoretical points of view
as tools for scientific simulations of complicated systems.
\end{abstract}

\Section{Introduction}

%
%
Scientific simulations have been widely used to study complicated
dynamic properties of physical and chemical systems along with
the recent development of high-performance computing environments.
However, when we investigate a large, complicated system,
which involves multiple scaled dynamics in space and time 
forming a dynamically hierarchical structure in the phase space,
the method of the direct numerical simulation (DNS) faces  
a serious obstacle. That is, relatively large resources are required
to achieve precise description of the whole behavior of the target system.
This is just the case for which we may introduce 
an alternative approach based on multi-physics simulations (MPS).
In MPS, 1) the target systems are divided into smaller parts
described by relatively simple physics; 2) each separated component  
is simulated independently according to the individually 
governing equation;
3) communication among separated components including
the dynamical interactions
is achieved by using ``coupled simulation techniques'', in
which a mediator plays a role of an information center for MPS
to control information among the extensive computational
resources.

Acoustic sound generation in musical instruments\cite{Fletcher1998}
is an interesting target of the numerical study,
for which a careful consideration 
in modeling physical processes as well as program codes
is required to  
to reproduce experimental results.
Many works have been published using empirical instrumental models
\cite{1996AdachiSato_Time,1981Schumacher,2000TachibanaTakahashi}
and it has been revealed that complicated dynamics
are induced by the coupling between airflows and generated sounds.
Recent studies on sound production
in air-reed instruments, e.g. flute, recorder, organ pipe and so on, 
often simulate 
jet oscillation by a resonator or turbulent dynamics with vortex shedding
using high-performance computing methods
\cite{Fletcher1998,2004Kuehnelt,2006Tsuchida}.
However, those approaches seem to achieve no more than partial success.
Thus, it is generally difficult to execute
DNS calculations for sound generation
since the calculations of compressible fluid dynamics in three dimensional
configurations are still too heavy in the usual computer environments.
One of the difficulties of the sound simulation
can be found in difference of the energy scale between turbulent flows
and radiated sounds, where the energy of the airflow
is $10^5$ times larger than the sound generated
under typical conditions in playing musical instruments.
It is practically impossible to study such problems of multi-scale 
properties by using the DNS.  Indeed, the larger
resouce is required to calculate detailed behaviors in the parts of the
smallest scales.
MPS or multiscale simulations (MSS) seem to provide
a breakthrough to solve those problems.
Furthermore, technical advantages of the MPS approach,
which is also called hybrid methods,
are generally known in computational aeroacoustics\cite{LESA2007}.

On the mechanism of the sound production by fast airflow around objects,
there have been many theoretical and numerical works
on the interaction between fluid dynamics and acoustic sounds\cite{LESA2007}.
The standard approach by the Lighthill's acoustic analogy\cite{1952Lighthill}
is widely used both in the recent reseaches based on the DNS strategy
and the approximate studies using Large-Eddy Simulations (LES)
of incompressible fluid dynamics.

In this paper, we apply the MPS technique to the
sound production by turbulent flows in musical instruments,
where a consistent coupling between the turbulent flow dynamics
and the sound propagation is considered.
This paper is organized as follows.
The basic theoretical and mathematical descriptions on the sound generation
are briefly reviewed in section \ref{sec:theory}.
Based on the theory, the coupled simulation system
of the turbulent flow and the sound propagation is constructed
in section \ref{sec:simulation}.
In section \ref{sec:cost},
computational costs for DNS and MPS approaches are evaluated,
and the concluding remark will be given in the last section.

\Section{Theories for sound generation}
\label{sec:theory}

The origin of sounds in flutelike instruments is usually considered
as the jet oscillations, vortex shedding,
and turbulences.  In this section, we briefly review the basic
theories on these sound production in musical instruments.

\SubSection{Lighthill's acoustic analogy}

The sound radiated from a finite region of turbulent flow is
estimated by Lighthill's theory on the acoustic analogy.
This provides a general approach to compute the field
where the acoustic pressure is generated by a function of the Lighthill
tensor\cite{2005Boersma-TCFD19-161,1952Lighthill,2001SSBJ-PhysF13-476}
\begin{equation}
  T_{ij}=\rho u_i u_j + (p-{c_0}^2\rho)\delta_{ij}-\sigma_{ij}.
\end{equation}
Under some assumptions,
it is known that only the main term of $T_{ij}$, i.e., $\rho u_i u_j$,
is necessary.
Then, the wave equation of sounds for the density of the fluid is
\begin{equation}
  \frac{\partial^2 \rho}{\partial t^2}
  -{c_0}^2\frac{\partial^2 \rho}{\partial x_i^2}
  =\frac{\partial^2T_{ij}}{\partial x_i\partial x_j}
\end{equation}

It is important that the Lighthill equation is an exact restatement
of the Navier-Stokes equations\cite{2005Boersma-TCFD19-161}.
This is the best starting point for the multiphysics approach
explained in the next section
since the mathematically exact representation is known.
However, it should be noted that
the validity of this description in the later time of the dynamics
is still nontrivial.
In the noise production simulations from turbulent flows,
the Lighthill's analogy seems to work well
\cite{2005Boersma-TCFD19-161,2001SSBJ-PhysF13-476}.
Since the noise propagates far from the source region
in the fluid immediately, complex effects with delayed or nonlinear
interactions seem to be small.

\SubSection{Nonlinear feedback in instruments}

In the musical instruments,
oscillating energy of the generated sound can feed back to the flows
after an interaction with the resonator.
This feedback effect plays the important role in the
sound generation in musical instruments
while it can make the problem complicated and difficult
\cite{1996AdachiSato_Trumpet,Fletcher1998,2000TachibanaTakahashi}.

In the usual approach to the dynamics in the instruments,
empirical models are used to simulate
the complicated nature from resonance and oscillation
\cite{Fletcher1998}.
In order to clarify the mechanism of the oscillation,
such a model approach often produces good results.
However, if we study the detailed theoretical mechanisms
in feedback interactions from a sound pressure to the fluids,
simulations with more realistic models should be used.
In the next, one of such approach using simulation techniques
will be presented.

\Section{Multiphysics simulation}
\label{sec:simulation}

\begin{figure}
\center\includegraphics[scale=0.5]{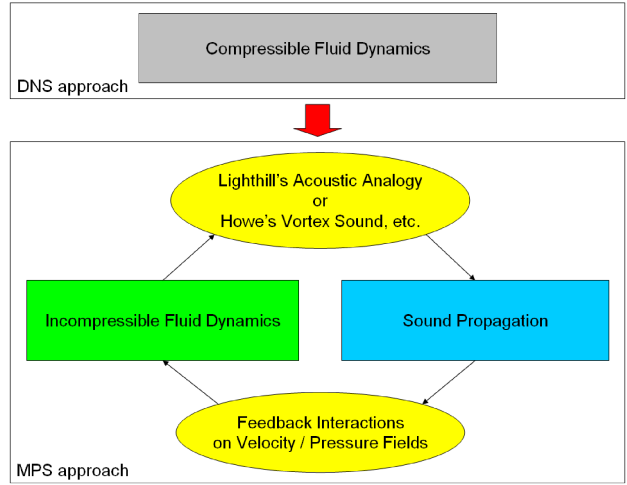}
\caption{Schematic picture of multiphysics simulation system for musical instruments}
\label{fig:schematic}
\end{figure}

Since the acoustic sound is an elastic wave on the air,
the sound generation by turbulent airflows in musical instruments
is fully simulated by the compressible fluid dynamics calculation
on detailed models of the instruments
with appropriate boundary and initial conditions.
If we require sufficient accuracy to the result
through the DNS calculation,
high-performance computing will be necessary.
The approach of the present work is opposite,
where active partitioning of the simulation system
into several components with different physical properties
is introduced.
This partitioning of the whole system is based on the physical
insight and the integration is performed by the MPS technique.

In the present problem, the sound generation in musical instruments,
the total system is divided into a part subject to
incompressible fluid dynamics by Navier-Stokes equations
and another part for sound propagation and radiation by wave equations.
These are calculated simultaneously with appropriate
coupling dynamics between fluid dynamics and acoustic waves.
The schematic picture of this total simulation system is given
in Fig.~\ref{fig:schematic}.

\begin{figure}[t]
\center\includegraphics[scale=0.22]{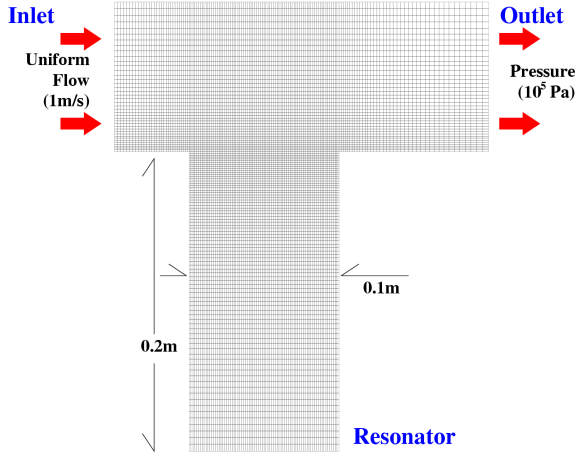}
\caption{Mesh structures of the inlet, outlet, and resonator.}
\label{fig:mesh}
\end{figure}

The model of the instrument in the present work
has the simplest box-resonator shape (Fig.~\ref{fig:mesh})
since the main subject of this work is to demonstrate
how useful the multiphysics approaches are.
The number of generated mesh is shown in Table~\ref{tbl:mesh}.
The boundary conditions of this simulation are following:
At the left inlet, a uniform flow with $1$ (m/s) is added,
and it runs through over the resonator to the right outlet
where the pressure is given $1$ bar ($=10^5$ Pa).
The upper wall is a non-friction wall with a velocity $1$ (m/s) as the fluid,
and at the other walls, the velocity is a constant $0$,
and the pressure is zero-gradient.
The mesh creation is done by {\tt blockMesh} utility
distributed within the open-source software package {\tt OpenFOAM-1.4}.
\footnote{OpenFOAM: {\tt http://www.opencfd.co.uk/openfoam/}}

\begin{table}
\caption{Mesh properties (two dimensional)}
\label{tbl:mesh}
\center\begin{tabular}{c|c|c}
\hline
\# Cells & \# Faces & \# Points\\ \hline
12400 & 25070 & 12671\\ \hline
\end{tabular}
\end{table}

\SubSection{Incompressible Large Eddy Simulation}

\begin{figure}
\center\includegraphics[scale=0.55]{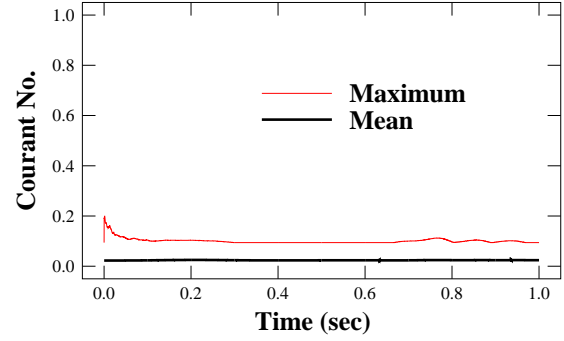}
\caption{Courant No. of the whole time evolution (1 second).}
\label{fig:CourantNo}
\end{figure}

Flows in the musical instruments
are considered as turbulent while they have
relatively low Mach and Reynolds numbers.
There are many numerical techniques to obtain time-evolution
of turbulent flows.  Recently, Large-Eddy Simulation (LES)
is introduced as a rather new tool in the field of aeroacoustics,
and is often used to simulate complicated fluid evolutions with turbulence.
In the present work, LES is used to calculate incompressible flows
in the instruments with appropriate thermo-physical parameters.
Although smaller scales are represented by the so-called subgrid models
in this method, it has been studied that acoustic sounds
generated by turbulent flows can be reproduced by the use of this model
\cite{2005Boersma-TCFD19-161,2001SSBJ-PhysF13-476}.
The LES equations using index notation and the summation convention
for spatially averaged field values are given by
\begin{equation}
\label{eq:LES}
  \frac{\partial \overline u_i}{\partial t}
  +\frac{\partial\overline u_i\overline u_j}{\partial x_j}
    =-\frac{\partial \overline p}{\partial x_i}
    +2\mu\frac{\partial \overline S_{ij}}{\partial x_j}
    -\frac{\partial\overline\tau_{ij}}{\partial x_j},
\end{equation}
\begin{equation}
  \frac{\partial \overline u_i}{\partial x_i}=0,
\end{equation}
where
$\overline u_j$ is a velocity field, $\rho_0$ is a constant density,
\begin{equation}
  S_{ij}\equiv\frac12\left(
    \frac{\partial \overline u_j}{\partial x_i}
    +\frac{\partial \overline u_i}{\partial x_j}
  \right)
\end{equation}
is a strain tensor, and
\begin{equation}
  \overline\tau_{ij}\equiv\overline{u_i u_j}-\overline u_i\overline u_j
  \label{eq:subgrid}
\end{equation}
is a subgrid scale tensor.

\begin{figure*}[t]
\begin{center}
\includegraphics[scale=0.2]{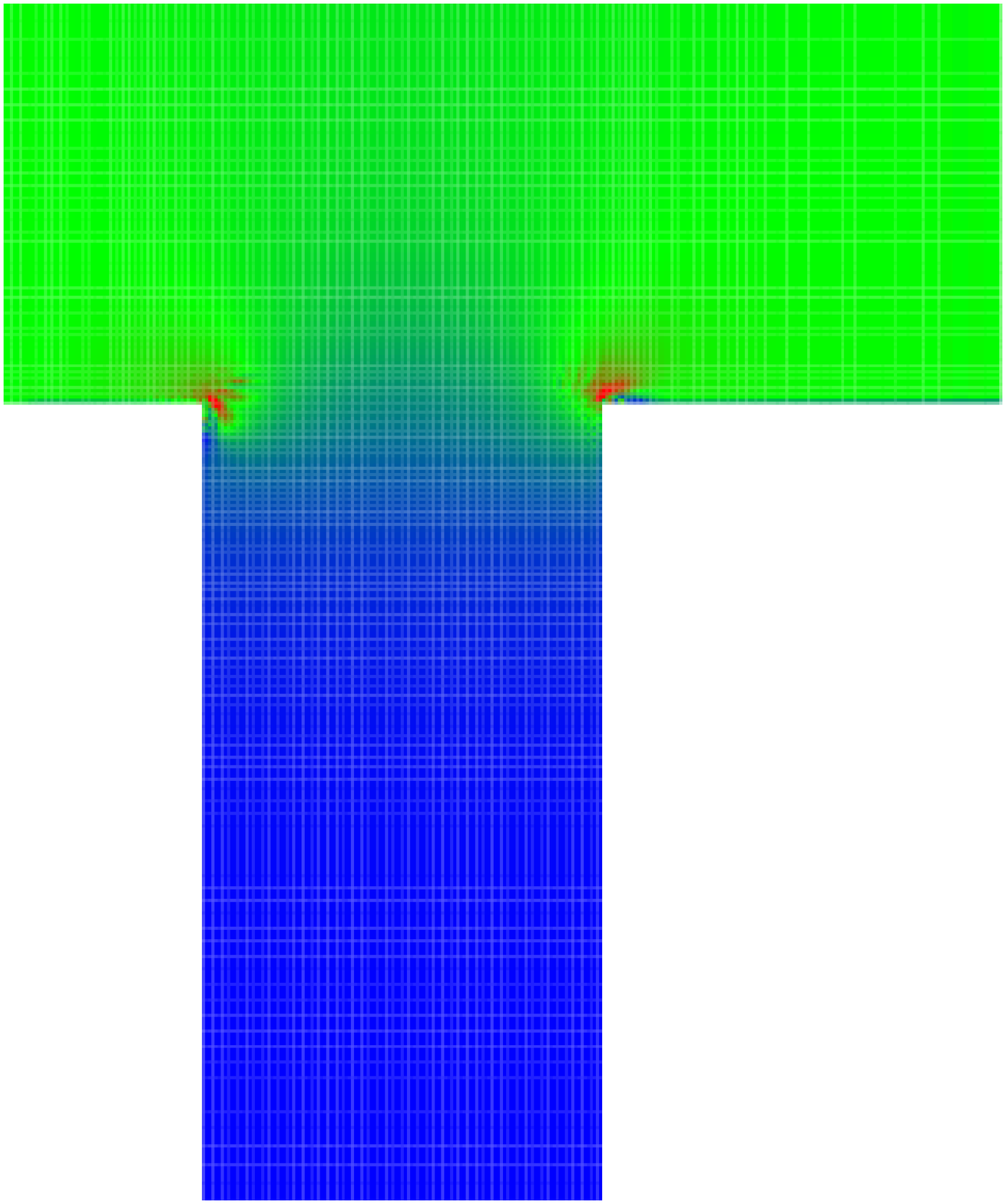}
\includegraphics[scale=0.2]{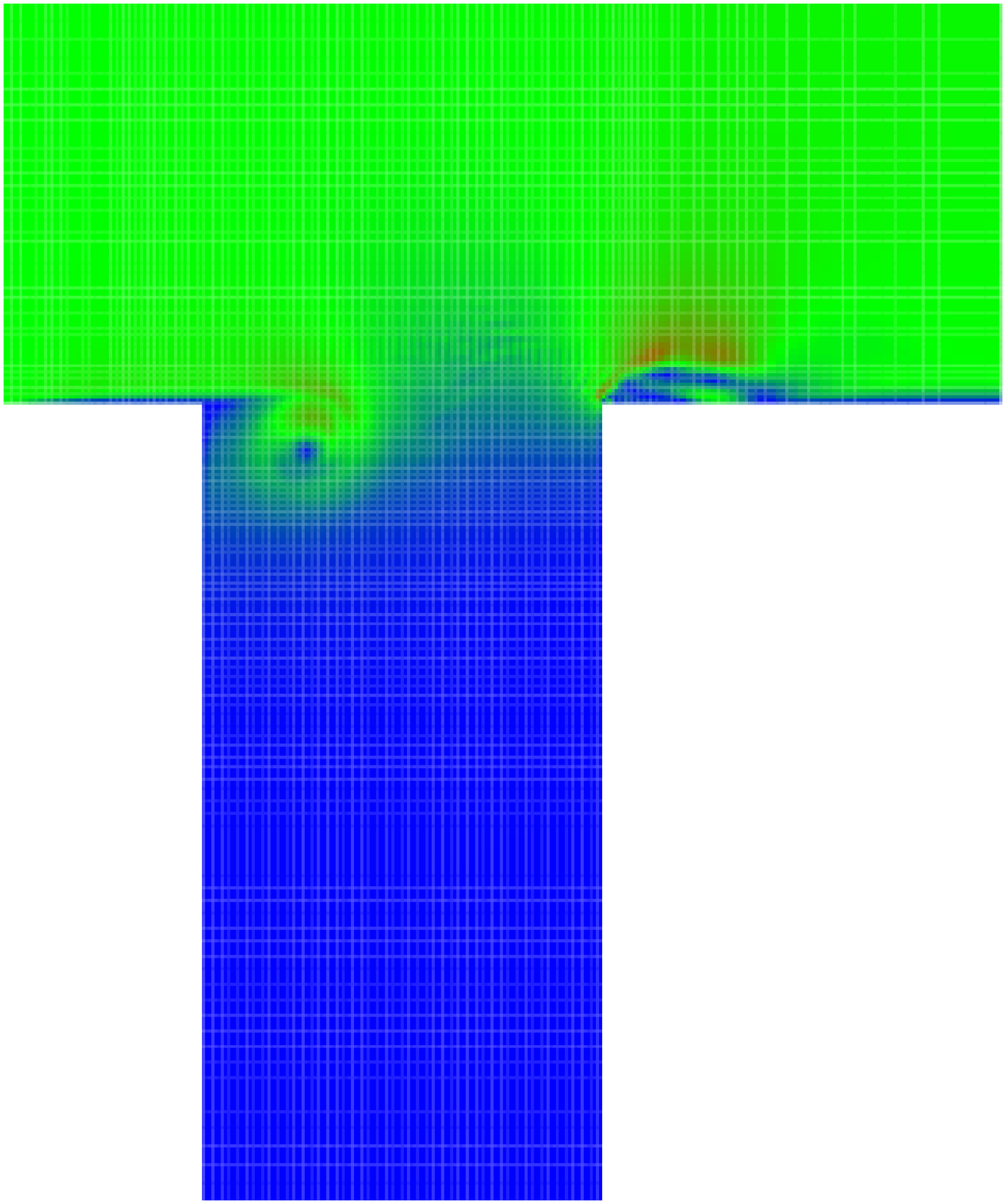}
\includegraphics[scale=0.2]{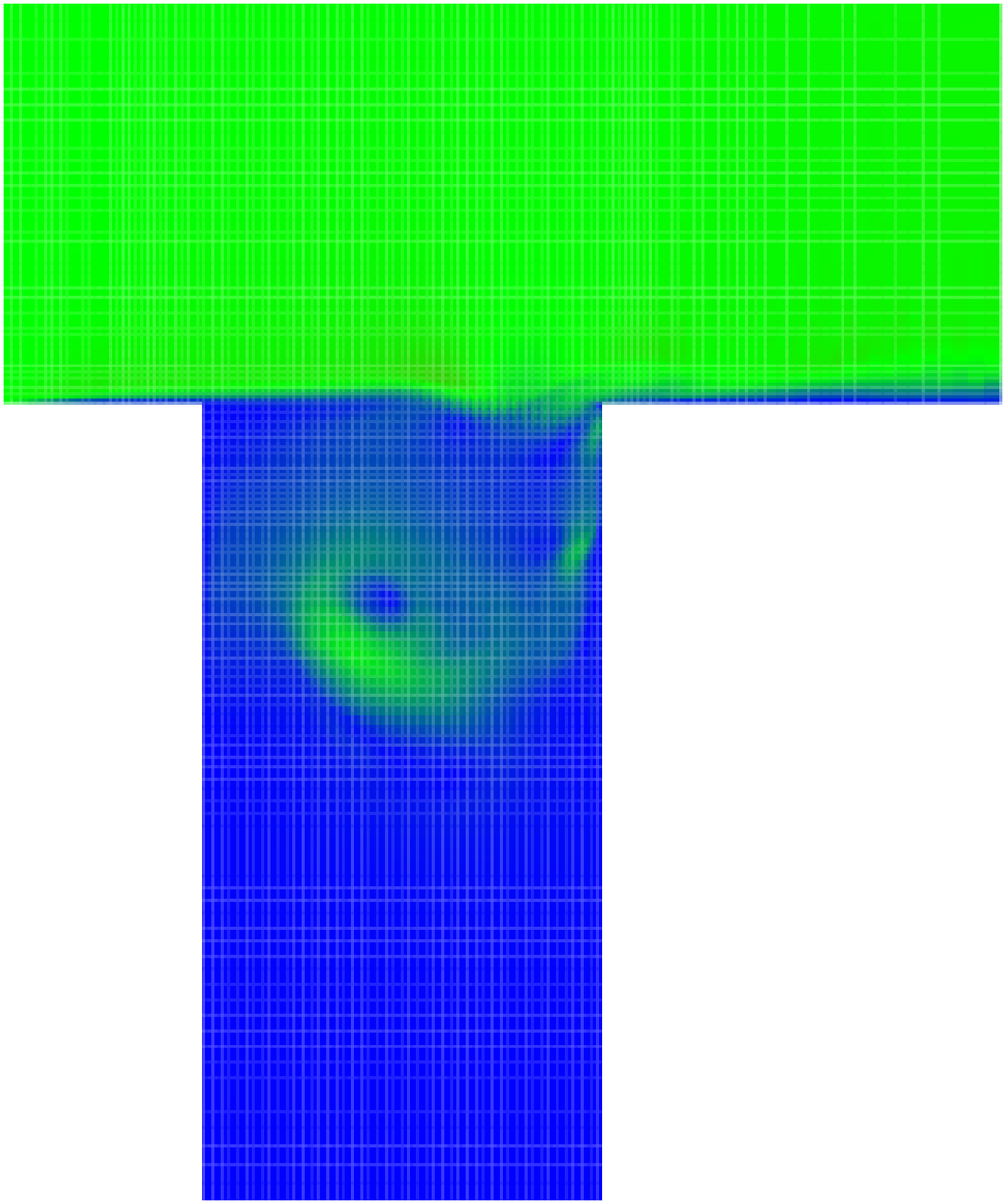}
\end{center}
\caption{Snapshot of incompressible fluid at t=0.01, 0.1, 1.0 (sec) (from left to right).}
\label{fig:U}
\end{figure*}

The computations are performed on a PC with Dual Xeon processors
since this is a test execution on the two dimensional minimum model.
The numerical scheme is based on the finite volume method (FVM)
and the time evolution scheme is the second order implicit method.
The total execution time was about three hours with $\Delta t=10^{-4}$ (sec).
The Courant numbers in the whole time evolution
are shown in Fig. \ref{fig:CourantNo}.
Snapshots of the results are presented in Fig. \ref{fig:U},
where the absolute values of the velocity $\overline u$
are shown at the times $t=0.01$, $t=0.1$, and $t=1.0$ (sec).

\SubSection{Sound production}

\begin{figure*}[t]
\begin{center}
\includegraphics[scale=0.2]{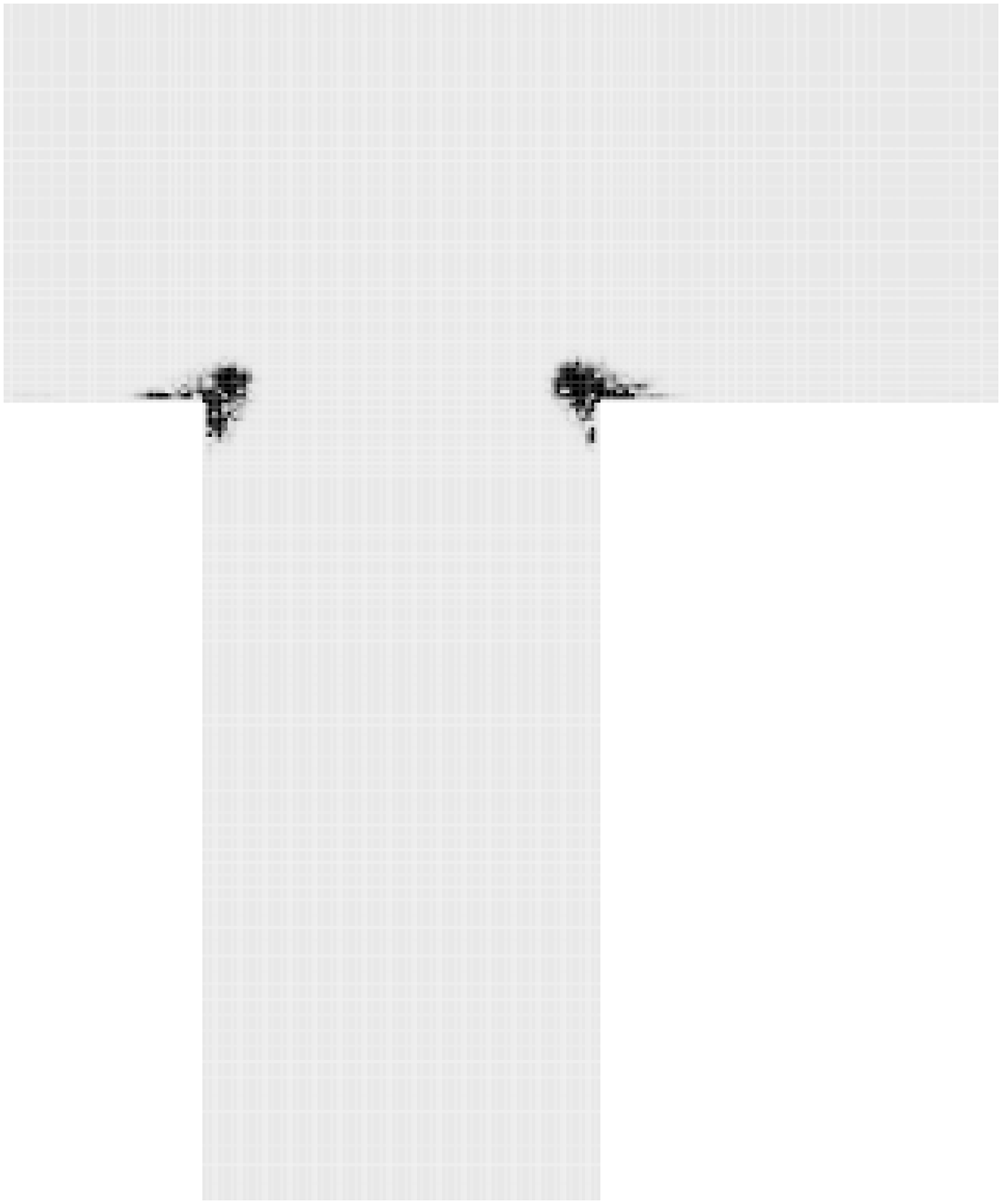}
\includegraphics[scale=0.2]{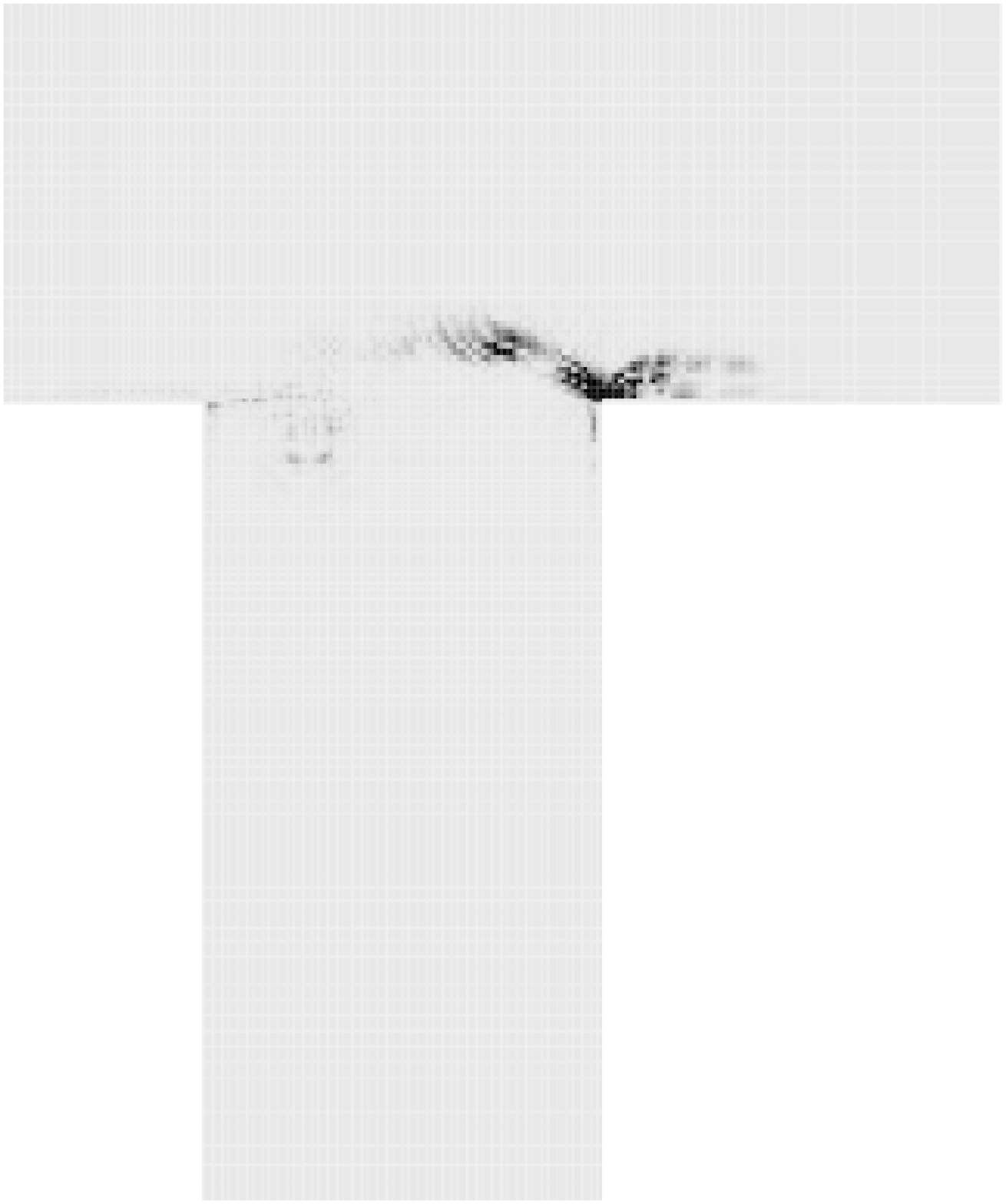}
\includegraphics[scale=0.2]{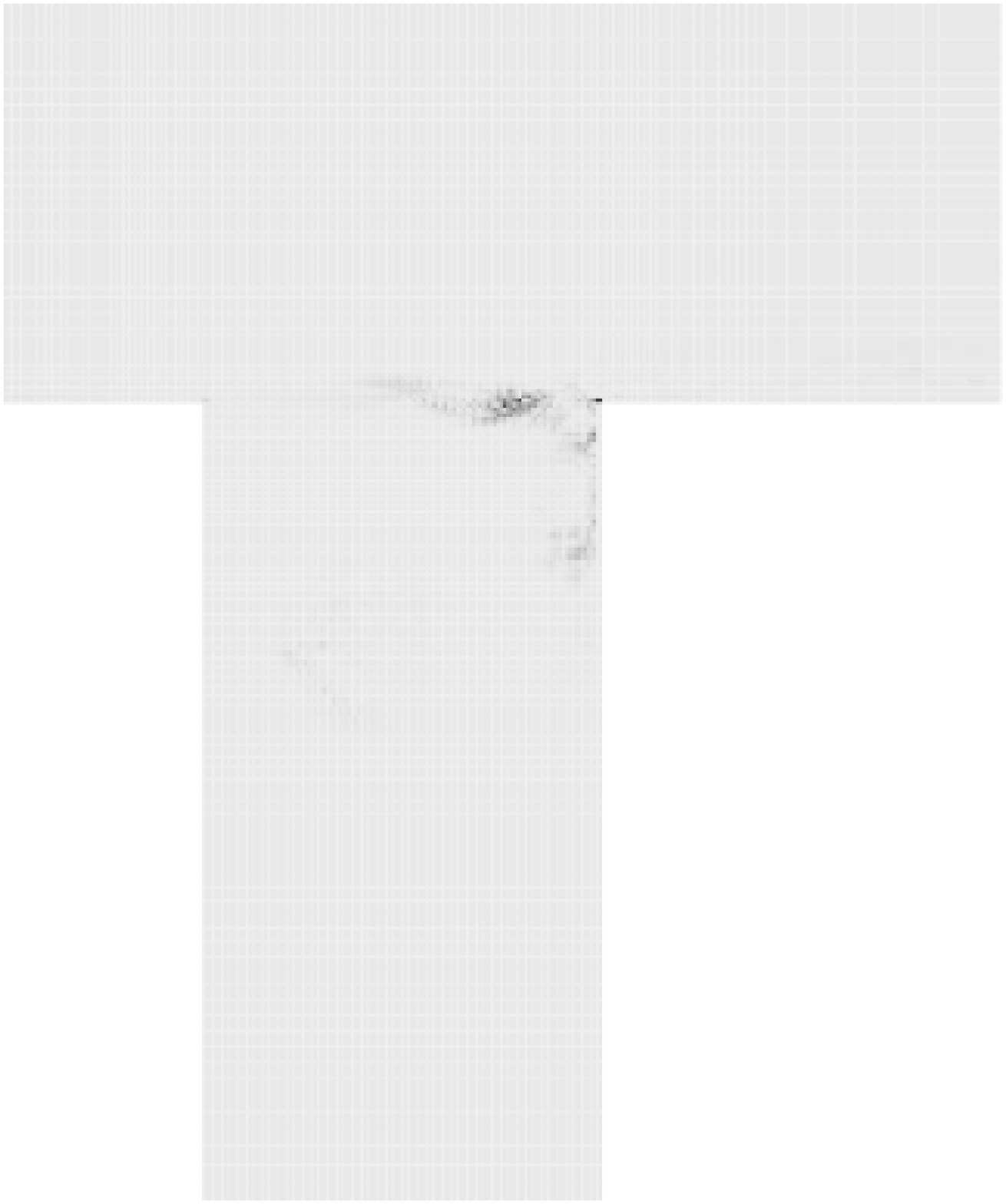}
\end{center}
\caption{Lighthill's acoustic source term at t=0.01, 0.1, 1.0 (sec) (from left to right).}
\label{fig:L}
\end{figure*}

Sound generation and propagation are represented in a wave equation
with Lighthill's source terms,
\begin{equation}
  \frac{\partial^2 \rho'}{\partial t^2}
  -{c_0}^2\frac{\partial^2 \rho'}{\partial{x_i}^2}
  =\frac{\partial^2 \overline T_{ij}}{\partial x_i\partial x_j}.
  \label{eq:Lighthill}
\end{equation}
where $\rho'$ represents a density by sound propagation, and
$T_{ij}$ is the Lighthill stress tensor which can be approximated by
\begin{equation}
  \overline T_{ij}=\rho_0\left(
    \overline u_i\overline u_j+\overline\tau_{ij}
  \right).
\end{equation}
This term can be obtained even in LES calculation
if we use the subgrid scale tensor, Eq. (\ref{eq:subgrid}).

In Fig.~\ref{fig:L},
the distribution of the Lighthill's source term,
the right-hand-side of Eq. (\ref{eq:Lighthill}),
is shown.
From the comparison with Fig. \ref{fig:U},
the sound sources are found around small areas where the flows are turbulent
and spatially non-uniform, i.e., near the corners or edges of walls.
In addition to these, we can find that vortices become the
sound source if they move, while it seems that a stable vortex
shown in the figure of $t=1.0$ does not produce sounds.

\SubSection{Closing the coupling}

In noise production simulations widely performed by DNS and LES,
the situation is rather simple since the noise is assumed to be spread out.
However, in musical instruments,
interactions between airflows and filtered sounds by a resonator
must be considered.
The sound is a density wave on the air as a compressible fluid
while the MPS system studied in the present work
is based on the incompressible fluid dynamics calculation.
In order to take into account the feedback effect from the sound
to the fluid, extra-terms must be introduced.
There are no standard and established strategies
on the feedback interaction so far.

We can present several forms to introduce the backward interaction.
If we assume that the air is ideal gas, we can use the equation
of state for an ideal gas,
\begin{equation}
  p=\rho RT
\end{equation}
where $R$ is the gas constant and $T$ is the temperature.
In the incompressible LES calculation, the density $\rho$ was
fixed $\rho=\rho_0$.
When we define small fluctuation term $\Delta\rho\equiv\rho'-\rho_0$
by the sound propagated, an additional term
\begin{equation}
  p\Rightarrow p+RT\Delta\rho
\end{equation}
can be introduced.
Then, this makes a closed loop to the velocity field
through the LES equation Eq. (\ref{eq:LES}).
What terms are necessary and sufficient should be evaluated through
these simulation studies.

\Section{Computational costs}
\label{sec:cost}

In this section, we estimate computational costs of
multiphysics simulations of musical instruments
which is compared to those of the direct numerical simulation.
So far in the present paper,
only two dimensional models are studied.
In this section, the cost will be estimated
for three dimensional models.

\SubSection{Memory requirements}

In our simulations described in the previous section,
LES calculation scheme is used where the stable results
can be obtained even if small scaled dynamics arises.
However, if we use DNS calculation for incompressible fluids
based on other numerical schemes,
finer mesh description is required for longer execution of the dynamics.
For two dimensional models, memory resources required may be
still small compared to the total amount of memory available
in the current computers.
For three dimensional models, the memory increase
by representing the fine structures of the compressible fluids
is rather critical and the amount of memory required
will be an order of 10GBytes.
This may not be impossible when we implement the parallel computation
over distributed memory machines,
but the performance of the program depends deeply
on the memory management.
Thus, the increase of the memory
makes longer the execution time of the total simulation.

\SubSection{Execution time}

When we use the MPS approach,
the mesh structures for the fluid dynamics and the sound propagation
can be defined separately.
The spatial mesh size $\Delta x$ for the fluid
is determined by the finest structure to be studied,
and $\Delta t\simeq \Delta x/U$ of the fluid dynamics
can be chosen rather large since the velocity $U$ of
the fluid is small ($\sim1$ (m/s) in instruments).
For representation of the sound propagation,
rather coarse mesh can be chosen since the spatial structure
is not so complicated around the basic frequency of the resonance.
On the other hand, time variable must be chosen rather small
because of the large velocity of the sound ($\sim$ 340m/s at
normal temperature).

If we perform DNS calculation for the same system,
a sufficiently fine representation for both the space and time must be chosen
so that the compressible fluid dynamics with the fast density
waves is properly simulated.  Then,
the computational time will be more than $10^2$ times longer
compared to the MPS approach.
In our two dimensional case, the CPU time for one step calculation
$\Delta t=10^{-4}$ is about 1 second.
It seems difficult to execute three dimensional DNS calculations
unless fine-tuned parallel programs are developed
and large-scale computer environments are available.

\Section{Conclusions}

\begin{figure}
\center\includegraphics[scale=0.45]{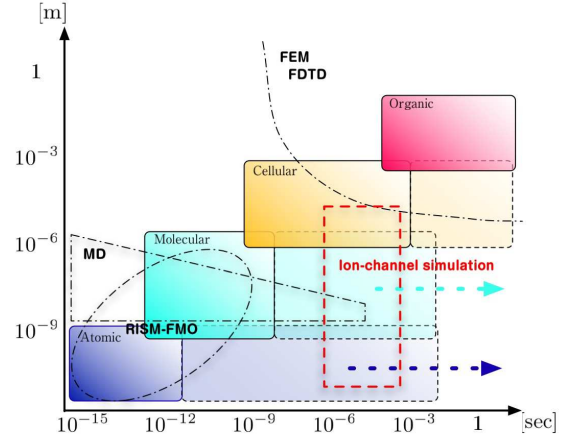}
\caption{Multiscale and multiphysics simulation describing the nature.}
\label{fig:multi}
\end{figure}

In the present work, acoustic sound production was simulated
using MPS approach which couples incompressible
fluid dynamics calculation by LES method
and sound propagation by wave equations.
It was apparent that the amount of computational costs
is rather small for the MPS case compared to the DNS approach,
where the separation of those programs was essential for systems
with different energy scales more than $10^5$.

This types of multiscale properties can be found in various
realistic systems (see Fig.~\ref{fig:multi}).
In the present system, incompressible Navier-Stokes equations
are established as basic equations to be solved,
and the DNS approach still works if we could provide
sufficient computational resources.
On the other hand, there are many cases that
it is difficult to introduce a
theory of description or a set of equations for the whole system,
when each component in the system has a different scale
or the total system is constructed from components
with originally different description\cite{TMO07}.
Description of the physics is hierarchical,
and a special approximation valid only to the scale is often used.
Then, it is difficult to determine which theory (or approximation) is basic.
Thus, multiphysics or multiscale approaches are significant,
and direct simulations can be useless even if sufficient
computational resources are available.

\bibliographystyle{latex8}
\bibliography{vs,MyWorks}

\begin{thebibliography}{10}\setlength{\itemsep}{-1ex}\small

\bibitem{1996AdachiSato_Time}
S.~Adachi and M.~Sato.
\newblock Time-domain simulation of sound production in the brass instrument.
\newblock {\em J. Acoust. Soc. Am.}, 99:1219--1226, 1996.

\bibitem{1996AdachiSato_Trumpet}
S.~Adachi and M.~Sato.
\newblock Trumpet sound simulation using a two-dimensional lip vibration model.
\newblock {\em J. Acoust. Soc. Am.}, 99:1200--1209, 1996.

\bibitem{2005Boersma-TCFD19-161}
B.~J. Boersma.
\newblock Large eddy simulatoin of the sound field of a round turbulent jet.
\newblock {\em Theoret. Comput. Fluid Dynamics}, 19:161--170, 2005.

\bibitem{Fletcher1998}
N.~H. Fletcher and T.~D. Rossing.
\newblock {\em The Physics of Musical Instruments}.
\newblock Springer-Verlag, New York, 2nd edition, 1998.

\bibitem{2004Kuehnelt}
H.~K\"uhnelt.
\newblock Simulating the sound generation in flutes and flue pipes with the
  lattice-bolzmann-method.
\newblock {\em Proceeding of ISMA}, pages 251--254, 2004.

\bibitem{1952Lighthill}
M.~J. Lighthill.
\newblock On sound generated aerodynamically.
\newblock {\em Proc. R. Soc. London Ser. A}, 211:564--587, 1952.

\bibitem{1981Schumacher}
R.~T. Schumacher.
\newblock Ab initio calculation of the oscillation in a clarinet.
\newblock {\em Acoustica}, 48:71--85, 1981.

\bibitem{2001SSBJ-PhysF13-476}
C.~Seror, P.~Sagaut, C.~bailly, and D.~Juv\'e.
\newblock On the radiated noise computed by large-eddy simulation.
\newblock {\em Physics of Fluids}, 13(2):476--487, 2001.

\bibitem{2000TachibanaTakahashi}
T.~Tachibana and K.~Takahashi.
\newblock Sounding mechanism of a cylindrical pipe fitted with a clarinet
  mouthpiece.
\newblock {\em Prog. Theor. Phys.}, 104:265--288, 2000.

\bibitem{TMO07}
T.~Takami, J.~Maki, J.~Ooba, T.~Kobayashi, R.~Nogita, and M.~Aoyagi.
\newblock Interaction and localization of one-electron orbitals in an organic
  molecule: Fictitious parameter analysis for multiphysics simulations.
\newblock {\em J. Phys. Soc. Jpn.}, 76(1):013001, 2007, arXiv: nlin.CD/0611042.

\bibitem{2006Tsuchida}
J.~Tsuchida, T.~Fujisawa, and G.~Yagawa.
\newblock Direct numerical simulation of aerodynamic sounds by a compressible
  cfd scheme with node-by-node finite elements.
\newblock {\em Comput. Methods Appl. Mech. Engrg.}, 195:1896--1910, 2006.

\bibitem{LESA2007}
C.~Wagner, T.~H{\"u}ttl, and P.~Sagaut, editors.
\newblock {\em Large-Eddy Simulation for Acoustics}.
\newblock Cambridge, 2007.

\end{thebibliography}

\end{document}